\documentclass[sigconf]{acmart}
\usepackage{graphicx} % Required for inserting images

\title{Reconsidering Conversational Norms in LLM Chatbots for Sustainable AI}

\author{Ronnie de Souza Santos}
\email{ronnie.desouzasantos@ucalgary.ca}
\orcid{}
\affiliation{%
  \institution{University of Calgary}
  \city{Calgary}
  \state{Alberta}
  \country{Canada}
}

\author{Cleyton Magalhaes}
\email{cleyton.vanut@ufrpe.br}
\orcid{}
\affiliation{%
  \institution{UFRPE}
  \city{Recife}
  \state{Pernambuco}
  \country{Brazil}}

\author{Italo Santos}
\email{isantos3@hawaii.edu}
\affiliation{%
  \institution{University of Hawai‘i at Mānoa}
  \city{Honolulu}
  \state{Hawaii}
  \country{USA}
  }

\begin{abstract}
LLM–based chatbots have become central interfaces in technical, educational, and analytical domains, supporting tasks such as code reasoning, problem solving, and information exploration. As these systems scale, sustainability concerns have intensified, with most assessments focusing on model architecture, hardware efficiency, and deployment infrastructure. However, existing mitigation efforts largely overlook how user interaction practices themselves shape the energy profile of LLM-based systems. In this vision paper, we argue that interaction-level behavior is an underexamined factor shaping the environmental impact of LLM-based systems, and we outline this issue across four dimensions. First, extended conversational patterns increase token production and raise the computational cost of inference. Second, expectations of instant responses limit opportunities for energy-aware scheduling and workload consolidation. Third, everyday user habits contribute to cumulative operational demand in ways that are rarely quantified. Fourth, the accumulation of context affects memory requirements and reduces the efficiency of long-running dialogues. Addressing these challenges requires rethinking how chatbot interactions are designed and conceptualized, and adopting perspectives that recognize sustainability as partly dependent on the conversational norms through which users engage with LLM-based systems.
\end{abstract}



\begin{document}

% Paper submission: https://icse2026-seis.hotcrp.com/paper/12/edit

% Opinion, vision, method, meta-research paper, up to 4 pages, reporting on well-founded arguments to support diversity and inclusion.
%
% For all papers, references may use 2 extra pages beyond the page limits stated above.

\maketitle

\section{Introduction}

\label{introduction}

Large Language Model (LLM)–based chatbots have become central to how users interact with computational systems across technical, educational, and analytical domains. In software engineering, conversational agents support developers as they reason about code, interpret design alternatives, and navigate complex information spaces through natural language dialogue~\cite{richards2024you,de2024unveiling}. Chat-based interfaces have also been adopted in educational and design-oriented environments, where they facilitate reflective learning, scaffold problem solving, and mediate interactions with analytical or development tools~\cite{richards2024you,ashkbous2025leveraging,becchi2025gpt,bekkar2024chatbots}. These developments indicate a shift toward LLM-driven conversation as a common mode of coordination between users and digital systems.

At the same time, sustainability concerns have gained prominence as research documents the resource demands of large-scale models. Studies examining training and inference describe how energy use is shaped by token processing, memory utilisation, and system-level behaviours~\cite{vartziotis2024carbon,jiang2024preventing,wilkins2024hybrid,stojkovic2403towards,ding2024sustainable}. %Broader sustainability frameworks emphasize that environmental impact depends on infrastructure, hardware efficiency, and patterns of use~\cite{van2021sustainable,falk2024challenging}. Existing mitigation strategies focus heavily on improving deployment infrastructure, scheduling workloads across heterogeneous hardware, and reducing emissions through operational decisions~\cite{wilkins2024hybrid}. Other work examined model reliability, since unnecessary generations and repeated corrections increase inference activity~\cite{jiang2024preventing}. These efforts aim to reduce the environmental footprint of LLMs, yet they address sustainability primarily through architectural and infrastructural interventions.
Broader sustainability frameworks emphasize that environmental impact arises not only from infrastructure and hardware efficiency but also from patterns of use~\cite{van2021sustainable,falk2024challenging}. Current mitigation strategies, however, focus predominantly on infrastructural and architectural levers, such as optimizing deployment environments, scheduling workloads across heterogeneous hardware, and reducing emissions through operational decisions~\cite{wilkins2024hybrid}. Related work has examined model reliability, noting that unnecessary generations and repeated corrections can inflate inference activity~\cite{jiang2024preventing}. Together, these efforts aim to reduce the environmental footprint of LLMs, yet they primarily address sustainability through system-level optimizations, giving less attention to how interaction practices contribute to overall impact.

A growing body of evidence suggests that conversational interaction itself may have a meaningful effect on the computational characteristics of LLM-based systems. Chatbots encourage extended exchanges, elaborated responses, and context-rich dialogue, all of which increase the number of processed tokens and shape memory usage during inference~\cite{richards2024you,de2024unveiling,coignion2024green,ashkbous2025leveraging}. Interaction norms such as immediate responsiveness and continuous multi-turn engagement reinforce real-time workloads that limit opportunities for energy-aware optimisation~\cite{wilkins2024hybrid}. As chat-based interfaces become integral to software engineering and other digital practices, understanding how user-facing behaviours influence the environmental profile of these systems becomes increasingly important~\cite{jiang2024preventing,wilkins2024hybrid}. This creates an opportunity to explore sustainable AI from the perspective of interaction design, complementing hardware- and model-level strategies with software-based considerations grounded in how users engage with LLM chatbots.

This vision paper examines four dimensions of this problem (Section~\ref{problem}). First, it analyses how conversational patterns influence the amount of generated text and contribute to the computational cost of inference. Second, it considers how expectations of immediacy limit opportunities for energy-aware scheduling and workload consolidation in real-time interaction. Third, it examines how user behaviour shapes the cumulative energy profile of chat-based systems over time. Fourth, it investigates how context accumulation increases memory requirements during inference and affects the efficiency of LLM-based chatbots. The paper concludes by outlining research directions for sustainable AI that focus on conversational norms and user-facing practices rather than hardware or infrastructure alone (Section~\ref{researchOpportunities}).
\section{Background}

This section provides background on software engineering bots, the sustainability concerns associated with LLMs, and the emerging paradox that arises when LLM-powered bots are deployed for an increasingly diverse range of purposes, including activities related to sustainability itself.

\subsection{Chatbots and LLMs in Software Engineering}

Chatbots have become established tools in software engineering, supporting both technical and social aspects of development work. Early studies describe chatbots as interfaces that assist developers by connecting them to services, providing feedback, and automating structured tasks across communication platforms~\cite{wessel2022software}. These systems participate in conversational channels, guide newcomers, surface relevant information, suggest code improvements, assist with defect investigation, and offer just-in-time explanations during development discussions~\cite{wessel2022software, moguel2023bots}.

With the introduction of LLMs, chatbots have become more capable and flexible. LLM-based chatbots support tasks such as code generation, bug explanations, summarization, and design reasoning, often within a single conversational interface~\cite{abedu2024llm}. Developers report that conversational interaction aligns with their problem-solving habits, enables incremental inquiry, and provides opportunities for learning through natural language explanations~\cite{ashkbous2025leveraging,richards2024you}. LLM-based chatbots thus extend earlier automation by offering adaptive dialogue and personalized guidance.

Although these systems enhance productivity and accessibility, they also introduce challenges, such as variable behaviour, inclusivity issues, context-handling limitations, and communication mismatches~\cite{richards2025bridging, melo2025enhancing}. As LLM-based chatbots increasingly support complex workflows, understanding how they operate and how they are used becomes important for assessing their broader implications. Additionally, because these chatbots rely on LLMs for their capabilities, their adoption intersects with growing concerns about the energy requirements and environmental impact of LLM-based systems~\cite{vartziotis2024carbon,jiang2024preventing,van2021sustainable}.

\subsection{LLMs and Sustainability}

Sustainability concerns have become central as LLMs scale in size, capability, and usage. Energy consumption in software systems has emerged as an important environmental and societal concern. Within AI, the concept of Green AI has been defined as research that produces novel results while explicitly accounting for computational cost and encouraging reductions in resource usage whenever feasible~\cite{schwartz2020green}. Applied to LLMs, this work has considered both training and inference impacts. Prior benchmarking studies have estimated the energy required to generate a single model response~\cite{samsi2023words}, and carbon analyses have compared the ongoing inference cost of different categories of machine learning systems, including task specific finetuned models and more general purpose models trained for multiple tasks. Deployment cost has been characterized as the energy and carbon required to perform a fixed number of inferences, such as 1,000 model outputs~\cite{luccioni2024power}.

Research investigating the environmental impact of AI systems indicates that large models require substantial energy for training, fine-tuning, and continuous inference~\cite{van2021sustainable,falk2024challenging,ding2024sustainable}. Carbon analyses of LLM services describe how operational costs depend on the number of processed tokens, hardware utilization, memory requirements, and the duration of system operation~\cite{vartziotis2024carbon,jiang2024preventing,wilkins2024hybrid}. According to Luccioni et al.~\cite{luccioni2024power}, inference may have an environmental impact comparable to model training, given the computational resources required to deploy modern models at scale. Although a single inference is far less costly than training, its higher frequency can lead to significant cumulative energy use. Inference is not cost-free, and the processing of each input–output sequence incurs energy expenditure that accumulates with repeated usage~\cite{falk2024challenging,jiang2024preventing}.

Studies of LLM inference show that longer input sequences require more computation, including increased memory access and processing time raises energy consumption relative to short-context queries~\cite{wilkins2024hybrid,jiang2024preventing,stojkovic2403towards}. These effects become particularly relevant in conversational settings where context grows with each turn. Broader sustainability frameworks in computing further note that the energy footprint of a system depends not only on architecture but also on patterns of use, including interaction frequency and runtime behaviors~\cite{van2021sustainable}. These observations highlight that LLM sustainability is influenced by architectural design, hardware efficiency, and user interaction patterns.

To address these issues, at the hardware level, some mitigation techniques have been explored. Research describes how heterogeneous allocation across energy-efficient and performance-oriented hardware can reduce resource use when tasks are scheduled according to workload size~\cite{wilkins2024hybrid,stojkovic2403towards}. Broader sustainability analyses highlight how data center characteristics, energy sources, and infrastructure decisions influence the environmental footprint of AI systems~\cite{van2021sustainable,falk2024challenging,ding2024sustainable}. Energy-aware scheduling and workload-adaptive allocation have also been shown to reduce energy consumption in LLM inference, although these improvements remain constrained by model size and interaction characteristics~\cite{wilkins2024hybrid,jiang2024preventing,stojkovic2403towards}. At the software level, fewer mitigation strategies have been discussed in the literature, suggesting that sustainability considerations remain concentrated primarily at the hardware and infrastructure layers rather than the interaction or application layers.

\subsection{The LLM–Sustainability Paradox}

LLM-based chatbots have been developed across multiple domains, reflecting their growing role as conversational interfaces for technical, educational, and analytical work. In software engineering, LLM-based chatbots help developers reason about code, understand design decisions, and obtain tailored explanations during problem-solving~\cite{richards2024you,ashkbous2025leveraging}. In educational contexts, conversational systems support learning and reflection through dialogue-based guidance, with applications ranging from programming instruction to environmental science education~\cite{zheng2025developing,nguyen2025value}. The energy footprint of LLM-based code assistants has been investigated through simulated developer interactions with GitHub Copilot, suggesting that energy consumption depends on factors such as model size, quantization, streaming, and concurrency, and that a substantial portion of generated suggestions is canceled or ignored, which introduces avoidable computation~\cite{coignion2024green}. The study further indicates that higher concurrency improves efficiency and that server configuration parameters, including GPU count and model size, influence energy use and latency, which points to practical opportunities for reducing environmental impact. Conversational systems have also been incorporated into sustainability-oriented applications, including eco design frameworks that assist engineers in environmentally informed decision making~\cite{ashkbous2025leveraging} and educational settings where chatbots support students in reasoning about climate change, ecological systems, and sustainability perspectives~\cite{nguyen2025value,bekkar2024chatbots}.

At the same time, studies examining the operational footprint of LLMs show that increases in input and output length elevate token processing, memory usage, and energy consumption during inference~\cite{wilkins2024hybrid,jiang2024preventing,vartziotis2024carbon,stojkovic2403towards}. In conversational settings, multi-turn exchanges, elaborated responses, and growing context windows further expand the amount of generated text and computational work~\cite{richards2024you,de2024unveiling,coignion2024green}. Sustainability frameworks emphasize that usage practices influence environmental impact, indicating that frequent or intensive interactions contribute to the overall energy profile of AI systems~\cite{van2021sustainable}. The paradox therefore emerges: while LLM-based chatbots are increasingly introduced to promote sustainability awareness and environmentally responsible decision making, their own operation relies on resource-intensive computation~\cite{ding2024sustainable}.
\section{Rethinking User Interaction in LLM-Based Chatbots}
\label{problem}

%In this vision paper, we explore open issues concerning the sustainability of LLM-based chatbots, with particular attention to how interaction design and user behaviour shape the environmental footprint of inference. The aim is not to prescribe solutions, but to characterize how conversational paradigms, expectations of immediacy, and context retention affect energy demand. Chatbots are positioned as intuitive tools for information access, yet their operational characteristics carry environmental implications that are often concealed from end users. We argue that user interaction should be viewed as part of the sustainability problem space, since interaction patterns influence token throughput, workload distribution, and memory usage across the inference lifecycle~\cite{stojkovic2403towards,ding2024sustainable}. The discussion below outlines four analytical dimensions that reveal how user-facing practices contribute to the computational demands of LLM-based chatbot systems.
This vision paper examines overlooked sustainability issues in LLM-based chatbots, focusing on how interaction design and user behaviour influence the environmental footprint of inference. Rather than prescribing solutions, we highlight how conversational para- digms, such as expectations of immediacy, extended dialogue, and persistent context shape computational demand in ways not typically accounted for. Although chatbots provide intuitive access to information, their operational characteristics carry environmental implications that remain largely invisible to end users. We argue that interaction practices are part of the sustainability problem space, affecting token throughput, workload management, and memory usage throughout the inference lifecycle~\cite{stojkovic2403towards,ding2024sustainable}. The following sections outline four analytical dimensions that illustrate how user-facing behaviour contributes to the resource demands of LLM-based systems.

\paragraph{\textbf{Interaction Patterns and the Cost of Output Inflation.}} Energy measurements increasingly indicate that the amount of generated text plays a substantial role in determining the computational cost of inference. Studies examining LLM inference pipelines describe how longer outputs require more computation per token, increase runtime, and reduce throughput across hardware systems~\cite{wilkins2024hybrid,stojkovic2403towards,coignion2024green}. Similar observations appear in carbon analyses of LLM-as-a-service workloads, where inference energy depends on the total number of processed tokens, including both prompt and generated output~\cite{vartziotis2024carbon}. Empirical evaluations of conversational assistants show that chatbots frequently produce elaborated or verbose responses, often exceeding what is necessary for user problem solving~\cite{richards2024you,de2024unveiling,ashkbous2025leveraging}. These findings suggest that extended responses, which are common in chatbot interactions, elevate operational demand even when a concise answer would suffice. The prevailing assumption that richer explanations inherently improve user experience therefore warrants reconsideration, particularly in contexts where shorter responses can adequately support user tasks.

\paragraph{\textbf{The Inefficiency of Real-Time Conversational Workloads.}} Chatbots are designed around immediate responsiveness, creating continuous, individualized workloads. However, research on energy-efficient inference indicates that meaningful optimization occurs when requests can be distributed across heterogeneous hardware or processed according to token thresholds~\cite{wilkins2024hybrid,stojkovic2403towards}. These strategies reduce energy use by allocating small workloads to more efficient systems, yet such allocation requires temporal flexibility that real-time chat typically cannot provide~\cite{jiang2024preventing,ding2024sustainable}. The emphasis on instantaneous response therefore limits opportunities for workload consolidation or deferred scheduling. Treating low-latency interaction as a fixed requirement obscures its environmental implications, and a more sustainable view would recognize that responsiveness is a design choice rather than an inherent constraint of conversational systems.

\paragraph{\textbf{User Behaviour and Demand-Side Sustainability.}} Sustainability concerns also emerge from user behaviour. Frameworks on sustainable software practice highlight that usage patterns, including the frequency and complexity of interactions, directly influence the energy profile of a system over time~\cite{van2021sustainable}. In LLM-based chatbots, routine look-ups, repeated small queries, and follow-up questions contribute cumulatively to token processing and operational energy use~\cite{jiang2024preventing}. Carbon analysis of LLM services similarly emphasizes that operational footprints are tied to the number of tokens processed during inference, indicating that elevated interaction volume increases downstream energy demand~\cite{vartziotis2024carbon}. Empirical studies of code assistants further show that unnecessary generations and unused suggestions increase energy consumption, and that manually triggered or selective invocation can reduce waste~\cite{coignion2024green}. These insights suggest that everyday interaction choices have measurable environmental effects. Introducing user-facing guidance or lightweight alternatives for simple tasks could support more sustainable patterns of use without reducing system utility.

\paragraph{\textbf{The Burden of Context Accumulation.}} Conversational agents commonly retain interaction histories to preserve coherence and continuity, which increases the number of tokens processed during inference. Existing evaluations of LLM inference show that energy consumption grows with the length of input sequences, as longer requests require additional computation and longer processing time~\cite{wilkins2024hybrid,jiang2024preventing,stojkovic2403towards}. Carbon analyses of LLM workloads similarly note that operational costs scale with the total number of processed tokens, indicating that accumulated dialogue history contributes to higher computational demand~\cite{vartziotis2024carbon}. Although retaining full conversational context can benefit certain tasks, maintaining long histories by default imposes additional energy use. Mechanisms such as selective summarization, shorter context windows, or user-controlled context persistence could reduce unnecessary computational overhead while preserving coherence where needed.

% \section{RETHINKING FAIRNESS TESTING}

% {\color{red}{
% What was the intended purpose of this section? Right now it looks like the content overlaps with section 3.2 (where the same notes have been copied). I suggest (1) we use this title as the title for section 3 and (2) we remove this section. Any thoughts?}}
% What would a dataset designed with another cultural perspective look like? 

% Localized to specific problems 

% It should have specific attributes related to cultural issues. 

% It could contain information related to the “cultural source” (e.g., NULL requires deletion in western culture, Buddhist culture: intentional emptiness as information) 

% It would need to capture knowledge systems that are not only western, but there could be challenge with gaining trust of underrepresented communities to harvest their data. While this may not address explainability challenges that currently plague AI systems, particularly LLMs, it would ensure representation of these communities.

\section{Research Opportunities} 
\label{researchOpportunities}

We identified research opportunities focused on how conversational interaction contributes to the environmental footprint of LLM-based chatbots and how new interaction models might support more sustainable patterns of use. These directions extend beyond hardware-centric mitigation and focus on user-facing, behavioral, and socio-environmental factors that shape the development of LLM-based chatbots.

\paragraph{\textbf{Quantifying and Mitigating Interaction Level Environmental Costs.}} Current research demonstrates that the total number of processed tokens strongly influences the environmental footprint of conversational systems~\cite{vartziotis2024carbon,wilkins2024hybrid,jiang2024preventing,stojkovic2403towards}. Standardized procedures for evaluating these factors remain limited. A research opportunity lies in developing metrics that characterize the cost of dialogue-level behaviors, including elaborated responses, multi-turn exchanges, and accumulated interaction history. Such work would support comparative analyses of interaction patterns and provide empirical grounding for sustainable design choices of LLM-based chatbots. This direction also connects to the need for strategies that manage conversational history more efficiently. While longer inputs require additional computation during inference~\cite{wilkins2024hybrid}, the contribution of accumulated dialogue to overall energy use remains insufficiently understood. Investigations into selective retention, summarization-based compression, or user-controlled context scope could show how different approaches influence both computational demand and user comprehension.

\paragraph{\textbf{Interaction and Task Models that Selectively Invoke Generative Reasoning.}} Many user queries do not require full generative reasoning. Routine look ups, factual retrieval, and simple transformations could be addressed through lightweight mechanisms such as templates, retrieval components, or smaller models. This suggests a need to classify tasks according to their computational requirements and identify thresholds for mode switching. Current interaction patterns often lead to repeated exchanges as users refine or narrow their requests~\cite{becchi2025gpt,coignion2024green}, which elevates token usage and contributes to operational energy cost~\cite{vartziotis2024carbon}. Research is needed on interaction models that reduce token throughput without diminishing utility. Possible directions include concise response modes, structured information presentation, and hybrid mechanisms that activate generative reasoning only when needed.

\paragraph{\textbf{Adaptive Invocation Mechanisms that Reduce Unnecessary Generations.}}
Coignion et al.~\cite{coignion2024green} show that a substantial portion of code-assistant generations are canceled or ignored, indicating that many requests do not warrant full model invocation. This points to the need for interaction and task models that trigger inference only when user intent is sufficiently clear. Predictive or interaction-aware invocation strategies, such as delaying generation until a query stabilizes or using lightweight intent-detection mechanisms could reduce unnecessary computation.

\paragraph{\textbf{User Guidance and Demand-Side Moderation.}} User behavior plays a core role in shaping cumulative energy consumption, as repeated inference and high interaction volume contribute directly to operational energy use~\cite{jiang2024preventing,coignion2024green}. Sustainability frameworks in computing observe that patterns of use contribute to long term environmental outcomes~\cite{van2021sustainable}. This creates opportunities to design interfaces that encourage low-impact practices, such as selecting minimal response modes for simple tasks or opting for lightweight alternatives when generative dialogue is unnecessary. Studying how users respond to such guidance, and how it influences their long-term interaction patterns, would support demand side strategies that complement infrastructural improvements.

\paragraph{\textbf{Sustainable Context Management.}} Conversational systems often retain interaction histories to preserve coherence, which increases the number of input tokens processed during inference. Existing studies show that energy consumption grows with input length and with the total number of tokens handled during inference~\cite{wilkins2024hybrid,jiang2024preventing,vartziotis2024carbon,stojkovic2403towards}. Research is therefore needed on managing conversational history in ways that balance coherence and efficiency. Potential directions include selective context retention, summarization-based compression, or user-controlled mechanisms for adjusting context scope. Studies on token-level environmental costs suggest that context-aware interaction design may reduce unnecessary computation, although the effects of different strategies on user comprehension and task performance remain insufficiently understood.

\paragraph{\textbf{Understanding Higher-Order and Systemic Effects.}} The environmental footprint of LLM-based chatbots extends beyond immediate computational costs. Large-scale deployments depend on data center infrastructure that affects water use, electronic waste, and the extraction of critical minerals~\cite{falk2024challenging,ding2024sustainable}. Interaction patterns define demand for this infrastructure, as repeated inference and sustained usage contribute to cumulative operational impact~\cite{jiang2024preventing}. Further research is needed to examine how increased chatbot adoption influences user behavior, induces rebound effects, or redirects tasks previously conducted through lighter tools toward generative systems. Evidence from sustainability-oriented chatbot deployments indicates that conversational systems can influence decision making~\cite{nguyen2025value,bekkar2024chatbots}, suggesting that indirect environmental consequences warrant further investigation. %\\

Overall, these research opportunities indicate that sustainable AI requires attention not only to hardware and infrastructure but also to interaction practices that shape inference workloads. By investigating task differentiation, response design, user behavior, context management, and systemic effects, future work can extend sustainability efforts into the software layer where users engage directly with LLM-based chatbots. However, these directions introduce trade-offs that warrant consideration, since reduced verbosity, relaxed immediacy, or constrained context may influence how users experience coherence and responsiveness. Interaction-level adjustments contribute to lower resource consumption, yet their effect remains bounded by infrastructural conditions that shape the environmental profile of LLM systems. A balanced view, therefore, recognizes both the value of conversational modifications and the practical limits imposed by large-scale model deployment.
\section{Conclusion} 
\label{conclusion}

This vision paper explored how the sustainability of LLM-based chatbots can be observed not only by aspects related to model architecture and deployment infrastructure but also by interaction design and user behavior. While ongoing work emphasizes improvements in hardware efficiency, deployment infrastructure, and system-level optimization, our analysis indicates that conversational practices also influence the environmental footprint of inference in meaningful ways. Extended responses, expectations of immediacy, and persistent context contribute to computational demand in ways that are often overlooked. The increasing use of LLM-based chatbots across software engineering activities amplifies the importance of these issues. As chatbots become integral to development workflows, the software engineering community will need sustainable perspectives for designing and deploying these systems.

Building on this motivation, we proposed research opportunities that foreground interaction-level considerations as part of sustainable AI. These include developing metrics for the environmental cost of dialogue behaviors, designing response models that reduce token demand, identifying which tasks require generative reasoning, supporting more sustainable user behaviors, and managing conversational history in ways that balance coherence and efficiency. These directions indicate that sustainability in LLM-based chatbot development requires attention to how users interact with chat-based systems and how software design choices influence long-term resource consumption.

\bibliographystyle{acm}
\bibliography{paper}

\end{document}